\documentclass[12pt,english]{article}
\usepackage[T1]{fontenc}
\usepackage[latin9]{inputenc}
\usepackage{amsmath}

\newcommand{\mnras}{MNRAS}

\newcommand{\apss}{APSS}
\newcommand{\jmp}{J. Math. Phys.}

\usepackage[svgnames]{xcolor}

\usepackage{babel}

\title{Metric of a Slow Rotating Body with Quadrupole Moment from the Erez-Rosen Metric}

\author{Francisco Frutos-Alfaro \\
Edwin Retana-Montenegro \\
Iv\'an Cordero-Garc\'ia, \\
Javier Bonatti-Gonz\'alez \\ 
{\small School of Physics, University of Costa Rica, San Pedro 11501, 
Costa Rica} \\ {\small frutos@fisica.ucr.ac.cr}}

\date{\today}

\begin{document}

\maketitle

\begin{abstract} 
A metric representing a slowly rotating object with quadrupole moment is 
obtained using a perturbation method to include rotation into the weak limit 
of the Erez-Rosen metric. This metric is intended to tackle relativistic 
astrometry and gravitational lensing problems in which a quadrupole moment has 
to be taken into account. 
\end{abstract}

\section{Introduction}

\noindent 
The first quadrupole solution to the Einstein field equations (EFE) was found 
by Erez \& Rosen (1959) \cite{Erez-Rosen}. Some errors were found in this 
derivation. These were later corrected by Doroshkevich {\it{et al}}. 
\cite{Doroshkevich1966}, Winicour {\it{et al}}. (1968) \cite{Winicor1968} and 
Young \& Coulter (1969) \cite{Young1969}. Other multipole solutions to the EFE 
were obtained by Quevedo (1986) \cite{Quevedo1986}, Quevedo (1989) 
\cite{Quevedo1989}, Quevedo \& Mashhoon \cite{Quevedo1991}, and 
Castej\'on {\it{et al}}. (1990) \cite{Castejon-Amenedo1990}. In the three first 
articles, the solutions were obtained with the help of the 
Hoenselaers-Kinnersley-Xanthopoulos (HKX) transformations 
\cite{Hoenselaers1979}, while in the latter, they used the Ernst formalism 
\cite{Ernst1968}. These authors obtain new metrics from a given seed metric. 
One can include other desirable characteristics (rotation, multipole moments, 
etc.) to these seed metrics by means of these formalisms. Recently, 
Boshkayev {\it{et al}}. (2012) \cite{Boshkayev} obtained an approximate 
solution describing the interior and exterior gravitational field of a slowly 
rotating and slightly deformed object.

\noindent 
The aim of this article is to derive an appropriate analytical tractable 
metric for calculations in astrometry and gravitational lens theory including 
the quadrupole moment and rotation in a natural form. For this new rotating 
metric, is not necessary a multipolar expansion in the potential to include 
the multipolar terms because the seed metric has already a quadrupole term, 
that is this metric is multipolar intrinsically.

\noindent 
This paper is organized as follows. In section \ref{sec:02}, we get the weak 
limit of the Erez-Rosen metric. The Lewis metric is presented in section 
\ref{sec:03}. The perturbation method is discussed in section \ref{sec:04}. 
The application of this method leads to a new solution to the EFE with 
quadrupole moment and rotation. It is checked by means of the 
REDUCE software \cite{Hearn1999} that the resulting metric is solution of the 
EFE. In section \ref{sec:05}, we compare our solution with the exterior 
Hartle-Thorne metric \cite{Hartle1968} in order to assure that our metric has 
astrophysical meaning. In section \ref{sec:06}, we transform the obtained 
metric using Cartesian coordinates. Forthcoming works with this metric are 
discussed in section \ref{sec:07}.

\section{Weak Limit of the Erez-Rosen Metric \label{sec:02}}

\noindent 
The Erez-Rosen metric \cite{Carmeli,Winicor1968,Young1969,Zeldovich} 
represents a body with quadrupole moment. The principal axis of the quadrupole 
moment is chosen along the spin axis, so that gravitational radiation can be 
ignored. This metric is given by

\begin{equation}
\label{erezros}
{d} s^{2} = {\rm e}^{2\psi} {d} t^{2} 
- {\rm e}^{2(\gamma-\psi)} {\tilde{\Delta}} \left(\frac{{d} r^2}{r^{2}-2 M r}
+ {d} \theta^{2} \right) 
- {\rm e}^{-2\psi} (r^{2}-2 M r) \sin^{2}{\theta} {d} \phi^{2} ,
\end{equation}

\noindent 
where $ M $ is the mass of the object and

\begin{equation}
\label{delta}
{\tilde{\Delta}} = r^{2}-2 M r + M^{2} \sin^{2}{\theta} ,
\end{equation}

\begin{eqnarray}
\label{psigam}
\psi & = & \frac{1}{2} 
\left\{\left(1 + \frac{q}{4} (3 \lambda^2 - 1)(3 \mu^2 - 1) \right) 
\ln{\left[\frac{\lambda - 1}{\lambda + 1}\right]} \right. \nonumber \\
& + & \left. \frac{3}{2} q \lambda (3 \mu^2 - 1) \right\} , \nonumber \\
\gamma & = & \frac{1}{2} (1+q)^2 
\ln{\left[\frac{\lambda^2 - 1}{\lambda^2 - \mu^2}\right]} \\
& - & \frac{3}{2} q (1 - \mu^2) \left[\lambda 
\ln{\left[\frac{\lambda - 1}{\lambda + 1}\right]} + 2 \right] \nonumber \\
& + & \frac{9}{4} q^2 (1 - \mu^2) \left[\frac{1}{16} ({\lambda^2 - 1}) 
(\lambda^2 + \mu^2 - 9 \lambda^2 \mu^2 - 1) 
\ln^2{\left[\frac{\lambda - 1}{\lambda + 1}\right]} \right. \nonumber \\
& + & \left. \frac{\lambda}{4} \left(\lambda^2 + 7 \mu^2 - 9 \lambda^2 \mu^2 
- \frac{5}{3} \right) \ln{\left[\frac{\lambda - 1}{\lambda + 1}\right]} 
\right. \nonumber \\
& + & \left. \frac{\lambda^2}{4} (1 - 9 \mu^2) 
+ \left(\mu^2 - \frac{1}{3}\right) \right] . \nonumber
\end{eqnarray}

\noindent
with $ \lambda = r / M - 1 $ and $ \mu= \cos{\theta} $. From now on, we will 
keep in the derivations terms up to order $ M^{2} $ and $ q M^{3} $. 
The approximate forms of $ \psi $ and $ \gamma $ are

\begin{equation}
\label{psi10}
\psi=\frac{1}{2}\ln{\left(1-\frac{2M}{r}\right)}
-\frac{2}{15} q \frac{M^{3}}{r^{3}} P_{2}(\cos{\theta}) 
+ O(M^3, \, q M^4, \, q^2),
\end{equation}
 
\begin{equation}
\label{gamma10}
\gamma=\frac{1}{2}\ln{\left(\frac{r^{2}-2 M r}{\tilde{\Delta}}\right)}
+ O(M^3, \, q M^4, \, q^2) ,
\end{equation}

\noindent 
where $ q = 15 G Q / (2 c^{2} M^{3}) $ with $ Q $ representing the quadrupole 
moment, and $ P_{2}(\cos{\theta}) = (3\cos^{2}{\theta} - 1)/2 $ is the second 
Legendre polynomial.

\noindent 
Defining the following variables

\begin{equation}
\label{calc}
{\cal C}:={\rm e}^{-\chi}
\simeq 1-\frac{2}{15} q \frac{M^{3}}{r^{3}} P_{2}(\cos{\theta}) 
+ O(M^3, \, q M^4, \, q^2) ,
\end{equation}

\begin{eqnarray}
\label{psi21}
{\cal F} & := & {\rm e}^{2\psi} =\left({1-\frac{2M}{r}}\right){\rm e}^{- 2 \chi} \\
& \simeq & {1-\frac{2M}{r}} 
- \frac{4}{15} q \frac{M^{3}}{r^{3}} P_{2}(\cos{\theta})
+ O(M^3, \, q M^4, \, q^2) , \nonumber
\end{eqnarray}

\noindent 
and

\begin{equation}
\label{gamma2}
{\cal G}:={\rm e}^{2\gamma} \simeq {\frac{r^{2}-2 M r}{\tilde{\Delta}}}
+ O(M^3, \, q M^4, \, q^2) .
\end{equation}

\noindent 
where

\begin{equation}
\label{chi}
\chi = \frac{2}{15} q \frac{M^{3}}{r^{3}} P_{2}(\cos{\theta}) .
\end{equation}

\noindent 
If we substitute the former definitions into \eqref{erezros}, the metric takes 
the form

\begin{eqnarray}
\label{aperezros}
{d} s^{2} & = & {\cal F} {d} t^{2} - \frac{1}{{\cal F}}\left[{d} r^{2}
+ r^{2} \left(1-\frac{2M}{r}\right) {d} \Sigma^{2}\right] \nonumber \\
& = & {\cal F} {d} t^{2} - \frac{{d} r^{2}}{{\cal F}} 
- {r^{2}}{\rm e}^{2 \chi} {d} \Sigma^{2} , 
\end{eqnarray}

\noindent 
where $ d\Sigma^{2} = d \theta^{2}+\sin^{2}{\theta}d\phi^{2} $, and the inverse of 
$ {\cal F} $ is written as

\begin{eqnarray}
\label{psi22}
\frac{1}{{\cal F}} & = & \left({1-\frac{2M}{r}}\right)^{-1} {\rm e}^{2 \chi} \\
& \simeq & {1+\frac{2M}{r}+\frac{4M^{2}}{r^{2}}}
+\frac{4}{15} q \frac{M^{3}}{r^{3}}P_{2}(\cos{\theta}) 
+ O(M^3, \, q M^2, \, q^2) . 
\nonumber 
\end{eqnarray}

\noindent 
It is interesting to note that the spherical symmetry is not presented in the 
weak limit.

\section{The Lewis Metrics \label{sec:03}}

\noindent 
The Lewis metric is given by \cite{Lewis,Carmeli} 

\begin{equation}
\label{lewis} 
{d}{s}^2 = V d t^2 - 2 W d t d \phi 
- {\rm e}^{\mu} d \rho^2 - {\rm e}^{\nu} d z^2 - X d \phi^2
\end{equation}

\noindent 
where we have chosen the canonical co\-or\-di\-na\-tes $ x^{1} = \rho $ and 
$ x^{2} = z $, $ V, \, W, \, X $, $ \mu $ and $ \nu $ are functions of $ \rho $ 
and $ z $ ($ \rho^2 = V X + W^2 $). Choosing $ \mu = \nu $ and performing the 
following changes of potentials

$$ V = f , \quad W = \omega f , \quad X = \frac{\rho^2}{f} - \omega^2 f 
\quad {\rm and} \quad {\rm e}^{\mu} = \frac{{\rm e}^{\gamma}}{f} , $$

\noindent 
we get the Papapetrou metric

\begin{equation}
\label{papapetrou} 
{d}{s}^2 = f (d t - 2 \omega d \phi)^2 
- \frac{{\rm e}^{\gamma}}{f} [d \rho^2 + d z^2] - \frac{\rho^2}{f} d \phi^2 .
\end{equation}

\section{Perturbing the Erez-Rosen Metric \label{sec:04}}

\noindent 
To include slow rotation into the Erez-Rosen metric we use the 
Lewis-Pa\-pa\-pe\-trou metric (\ref{papapetrou}). First of all, we choose 
expressions for the canonical coordinates $ \rho $ and $ z $. For the Kerr 
metric \cite{Kerr63}, one particular choice is \cite{Carmeli,Chandrasekhar} 

\begin{equation}
\label{chandra} 
\rho = \sqrt{\Delta} \sin{\theta} \qquad {\rm and} \qquad 
z = (r - M) \cos{\theta} 
\end{equation}

\noindent 
where $ \Delta = r^2 - 2 M r + a^2 \simeq r^2 - 2 M r = r^2 {\cal F} e^{2 \chi} $.

\noindent 
From (\ref{chandra}) we get

\begin{eqnarray}
\label{cylindric}
d \rho^2 + d z^2 & = &
[ {(r - M)^2} \sin^2{\theta} + \Delta \cos^2{\theta} ] 
\left( \frac{d r^2}{\Delta} + d {\theta}^2 \right) \\
& \simeq & \left(1 + \frac{M^2}{r^2} \sin^2{\theta} \right) d r^2
+ r^2 \left(1 - \frac{2 M}{r} + \frac{M^2}{r^2} \sin^2{\theta} \right) 
d {\theta}^2 \nonumber 
\end{eqnarray}

\noindent 
where we have expanded up to $ M^2 $ order.

\noindent 
Now, let us choose $ V = f = {\cal F} $ and neglect the second order in 
$ \omega $ ($ \omega^2 \simeq 0 \Rightarrow W^2 \simeq 0 $). Then, we have

$$ X \simeq {\rho^2}{f} {r^2}{\rm e}^{2 \chi} \sin^{2}{\theta} . $$

\noindent 
If we choose 

$$ {{\rm e}^{\mu}} = \frac{r^2 {\rm e}^{2 \chi}}{{(r - M)^2} \sin^2{\theta} 
+ \Delta \cos^2{\theta}} , $$
 
\noindent 
the term (\ref{cylindric}) becomes

$$ {\rm e}^{\mu} [d \rho^2 + d z^2] = \frac{{d} r^{2}}{{\cal F}} 
+ {r^{2}}{\rm e}^{2 \chi} {d} {\theta}^{2} . $$

\noindent 
This term appears in the approximate Erez-Rosen metric (\ref{aperezros}). 

\noindent 
From (\ref{papapetrou}), let us propose the following metric 

\begin{equation}
\label{lewis2} 
{d}{s}^2 = V d t^2 - 2 W d t d \phi - Z {{d} r^{2}} 
- Y {d} \theta^{2} - X {d} \phi^{2} , 
\end{equation}

\noindent 
where $ X = {r^{2}}{\rm e}^{2 \chi} \sin^{2}{\theta} $, 
$ Y = {r^{2}}{\rm e}^{2 \chi} $, and $ Z = 1 / V $.

\noindent 
We see that to obtain a slowly rotating version of the metric (\ref{aperezros}) 
the only potential we have to find is $ W $. Then, the EFE must be solved:

\begin{equation}
\label{einstein} 
G_{i j} = R_{i j} - \frac{R}{2} g_{i j} = 0 
\end{equation}

\noindent 
where $ R_{i j} $ ($ i, \, j = 0, \, 1,\, 2, \, 3 $) are the Ricci tensor 
components and $ R $ is the curvature scalar. 

\noindent 
Fortunately, the Ricci tensor components 
$ R_{0 0}, \, R_{1 1}, \, R_{1 2}, \, R_{2 2}, \, R_{2 3}, \, R_{3 3} $ and the 
curvature scalar $ R $ depend upon the potentials $ V, \, X, \, Y, \, Z $ and 
not on $ W $. Therefore, these components vanish (see appendix). The only 
remaining equation we have to solve is $ R_{0 3} = 0 $, because it depends upon 
$ W $. The equation for this component up to the order 
$ O(M^3, \, a^2, \, q M^4, \, q^2) $ is

\begin{equation}
\label{eqdif} 
2 (1 - P_{2} (\cos{\theta})) \left[\frac{\partial^2 W}{\partial \theta^2} 
+ r^2 \frac{\partial^2 W}{\partial r^2} \right] 
- 3 \cos{\theta} \sin{\theta} \frac{\partial W}{\partial \theta} = 0 .
\end{equation}

\noindent 
The solution for this equation is

\begin{equation}
\label{soleqdif}
W = \frac{\cal K}{r} \sin^{2}{\theta} .
\end{equation}

\noindent 
where $ {\cal K} $ is a constant that we have to find. This constant can be 
found from the Lense-Thirring metric which can be obtained from the Kerr 
metric, {\it i.e.}

\begin{equation}
\label{lense-thirring}
{d} s^{2}= \left(1 - \frac{2 M}{r} \right) {d} t^{2} 
+ \frac{4 J}{r}\sin^{2}{\theta}{d} t {d}\phi
- {\left(1 - \frac{2 M}{r} \right)^{- 1}} {{d} r^{2}} - {r^{2}}{d} \Sigma^{2} ,
\end{equation}

\noindent 
where $ J = M a $ is the angular momentum and $ a $ is the rotation parameter. 

\noindent 
Comparing the second term of the latter metric with the corresponding of the 
metric (\ref{lewis2}), {\it i.e.} $ W $, we note that 
$ {\cal K} = - 2 J = - 2 M a $.

\noindent 
Then, the new rotating metric with quadrupole moment written in standard form 
\cite{Weinberg1972} in spherical coordinates is

\begin{eqnarray}
\label{newerezros}
{d} s^{2} & = & \left({1-\frac{2M}{r}}\right) {\rm e}^{-2 \chi} {d} t^{2} 
+ \frac{4 J}{r}\sin^{2}{\theta}{d} t {d}\phi
- \left({1-\frac{2M}{r}}\right)^{-1} {\rm e}^{2 \chi} {{d} r^{2}} \nonumber \\
& - & {r^{2}}{\rm e}^{2 \chi} ({d} {\theta}^{2} + \sin^2{\theta} {d} {\phi}^{2}) . 
\end{eqnarray}

\noindent 
We verified that the metric \eqref{newerezros} is indeed a solution of the EFE 
using REDUCE \cite{Hearn1999} up to the order $ O(a^2, \, q M^4, \, q^2) $. 
Hence, one does not need to expand the term $ ({1-{2M}/{r}})^{-1} $ in a Taylor 
series. 

\section{Comparison with the Exterior Hartle-Thorne Metric \label{sec:05}}

\noindent 
In order to establish whether our metric does really represent 
the gravitational field of an astrophysical object, one should show that it is 
possible to construct an interior solution, which can appropriately be matched 
with the exterior solution. For this purpose, we employ the exterior 
Hartle-Thorne metric \cite{Hartle1968,Berti}, which is given by

\begin{eqnarray}
\label{hartle}
d s^2 & = & 
\left(1 - \frac{2 {\cal M}}{r} 
+ \frac{2 {\cal Q} {\cal M}^3}{r^3} P_2(\cos{\theta}) \right) d t^2 \nonumber \\
& - & \left(1 + \frac{2 {\cal M}}{r} + \frac{4 {\cal M}^2}{r^2} 
- \frac{2 {\cal Q} {\cal M}^3 }{r^3} P_2(\cos{\theta}) \right) d r^2 \\
& - & r^2 \left(1 - \frac{2 {\cal Q} {\cal M}^3}{r^3} P_2(\cos{\theta}) \right) 
d \Sigma^2 + \frac{4 {\cal J}}{r} \sin^2{\theta} d t d \phi , 
\nonumber  
\end{eqnarray}

\noindent 
where $ {\cal M} $, $ {\cal J} $, and $ {\cal Q} $ are related with the total 
mass, angular momentum, and mass quadrupole moment of the rotating object, 
respectively.

\noindent 
Now, comparing the exterior Hartle-Thorne metric with our expression 
(\ref{newerezros}), it can easily be seen that upon defining 

\begin{equation}
\label{definitions}
{\cal M} = M, \qquad {\cal J} = J, \qquad 
2 {\cal Q} {\cal M}^3 = - \frac{4}{15} q {M^3} , 
\end{equation}

\noindent 
both metrics coincide up to the order $ O(M^3, \, a^2, \, q M^4, \, q^2) $. 
Our approximate expression for the Hartle-Thorne metric (\ref{hartle}) was 
obtained by means of a REDUCE program using the expressions from Abramowicz 
{\it et al.} \cite{Abramowicz}. We compared these results with the approximate 
expression given by Boshkayev {\it et al.} \cite{Boshkayev} and found that 
they have an extra term of order $ O (q M^4) $, which we neglected, 
because it is beyond the order we are working with. Additional differences are 
that our metric parameters ($ M, \, J = M a, \, q M^3 $) are distinct and our 
expressions (\ref{definitions}) are simpler than those of 
Boshkayev {\it et al.} \cite{Boshkayev}.

\section{The Transformation of the Metric \label{sec:06}}

\noindent 
In some cases, the metric \eqref{newerezros} has to be transform from spherical 
$(r, \, \theta, \, \phi)$ into Cartesian coordinates $(x, \, y, \, z)$. 
For example, if a comparison with a post-Newtonian (PN) metric is made, 
we have to transform the metric \eqref{newerezros} by using one of the 
following radial coordinates transformation: the harmonic or the isotropic 
coordinates of Schwarzschild metric. The first one is $r={\bar{r}}+M$, and 
the second one is $r={\bar{r}}(1+{M}/{2{\bar{r}}})^{2}$, where ${\bar{r}}$ is 
a new radial coordinate \cite{Weinberg1972}. We choose the first one, then the 
metric \eqref{newerezros} is transformed into

\begin{equation}
\label{newerezros2}
{d} s^{2} = {\cal H}{d} t^{2} + \frac{4J}{\bar{r}}\sin^{2}{\theta}{d} t {d} \phi
- \frac{{d} {\bar{r}}^{2}}{{\cal H}} - {{\bar{r}}^{2}}{\rm e}^{2 \chi}
\left(1+\frac{M}{\bar{r}}\right)^2 {d} \Sigma^{2} , 
\end{equation}

\noindent 
where

\begin{eqnarray}
\label{calh1}
{\cal H} & = & \left(\frac{{1-\frac{M}{\bar{r}}}}{{1+\frac{M}{\bar{r}}}}\right)
{\rm e}^{- 2 \chi} \\ 
& \simeq & {1-\frac{2M}{\bar{r}}+\frac{2M^{2}}{{\bar{r}}^{2}}} 
-\frac{4}{15} q \frac{M^{3}}{{\bar{r}}^{3}} P_{2}(\cos{\theta})
+ O (M^3, \, a^2, \, q M^4, \, q^2) , \nonumber 
\end{eqnarray}

\noindent 
and

\begin{eqnarray}
\label{calh2}
\frac{1}{{\cal H}} & = & \left(\frac{{1+\frac{M}{\bar{r}}}}
{{1-\frac{M}{\bar{r}}}}\right) {\rm e}^{2 \chi} \\ 
& \simeq & {1+\frac{2M}{\bar{r}}+\frac{2M^{2}}{{\bar{r}}^{2}}} 
+\frac{4}{15} q \frac{M^{3}}{{\bar{r}}^{3}} P_{2}(\cos{\theta})
+ O(M^3, \, a^2, \, q M^4, \, q^2) . \nonumber
\end{eqnarray}

\noindent 
Noting that $ {\cal C} = {\rm e}^{\chi} $ is still given by \eqref{calc} with 
$ r $ changed by $ {\bar{r}} $. Now, dropping the bar on $ r $, we transform 
the metric \eqref{newerezros2} into the Cartesian coordinates which are given 
by the usual relations

\begin{eqnarray}
x & = & r \sin{\theta}\cos{\phi} , \nonumber \\
y & = & r \sin{\theta}\sin{\phi} , \nonumber \\
z & = & r \cos{\theta}.
\end{eqnarray}

\noindent 
The resulting metric has the following form

\begin{equation}
\label{newerezros3}
{d} s^{2}={\cal H}{d} t^{2} + \frac{4J}{r^{3}} {d} t (x {d} y-y {d} x)
-\frac{\cal P}{r^{2}} ({\bf x} \cdot {d}{\bf x})^{2}
- \left(1+\frac{M}{r}\right)^2 {\rm e}^{2 \chi} {d}{\bf x}^{2} , 
\end{equation}

\noindent 
where 

\begin{eqnarray}
\label{var}
{\cal P} &=& \frac{1}{{\cal H}}
- \left(1+\frac{M}{r}\right)^2 {\rm e}^{2 \chi} \\ 
& \simeq & \frac{M^2}{r^2} + O(M^3, \, a^2, \, q M^4, \, q^2) . \nonumber
\end{eqnarray}

\noindent 
The former metric can be generalized as follows

\begin{equation}
\label{newerezros4}
{d} s^{2}={\cal H}{d} t^{2} + 8 {\bf V} \cdot {d}{\bf x}{d} t 
-\frac{\cal P}{r^{2}}({\bf x} \cdot {d}{\bf x})^{2} 
- \left(1+\frac{M}{r}\right)^2 {\rm e}^{2 \chi} {d}{\bf x}^{2} , 
\end{equation}

\noindent 
where 

\begin{eqnarray}
\label{var2}
\left(1+\frac{M}{r}\right)^2 {\rm e}^{2 \chi} & \simeq & 1+\frac{2 M}{r}
+\frac{M^2}{r^2}
+ \frac{4}{15} q \frac{M^{3}}{{\bar{r}}^{3}} P_{2}(\cos{\theta}) \nonumber \\ 
& + & O(M^3, \, a^2, \, q M^4, \, q^2) , 
\end{eqnarray}

\noindent
and the vector ${\bf V}$ is defined as

\begin{equation}
\label{vectorv}
{\bf V}=\frac{G}{2c^{3}r^{3}}\left[{\bf J}\times{\bf x}\right] , 
\end{equation}

\noindent 
with ${\bf J}=J {\hat{{\bf e}}_{J}}$ (${\hat{{\bf e}}_{J}}$ an unit vector in 
the direction of ${\bf J}$. The expressions for ${\cal H}$ and ${\cal H}^{-1}$ 
are valid changing ${\bar{r}}$ by $r$ in eqs. \eqref{calh1} and \eqref{calh2}.

\section{Conclusion \label{sec:07}}

\noindent 
In this paper, we include the rotational effect using the weak limit of the 
Erez-Rosen metric as seed metric into the Lewis-Pa\-pa\-pe\-trou metric. 
Thus, a new metric with quadrupole moment and rotation in the 
weak limit is obtained. Generally speaking, the quadrupole moment is included 
in the metric, for instance, in gravitational lensing, through the expansion of 
the gravitational potential in a power series \cite{Asada2005}. The resulting 
metric from our calculations includes the quadrupole moment in a natural form 
and is similar to the exterior metric obtained by Boshkayev {\it {et al}}. 
\cite{Boshkayev}. 

\noindent 
As we have seen in section \ref{sec:04}, our new metric agrees with the 
Hartle-Thorne solution \cite{Hartle1968}, whom obtained an interior metric that 
appropriately matches their exterior one, which guarantees the construction of 
an interior solution for our spacetime. This result indicates that our metric 
may be used to represent a compact astrophysical object.   

\noindent 
The new metric has many applications. For example, in calculations involving 
relativistic astrometry, in gravitational lens theory or planetary perihelion 
shift, it is useful to have a metric that include rotation and quadrupole 
moment. In relativistic astrometry, one needs a post-Newtonian metric to get 
after some approximations the deflection angle. It allows to get 
expressions for the right ascension $ \alpha $ and declination $ \delta $ for a 
celestial body \cite{Soffel1989,Soffel1985} in the gravitational field 
including rotation and quadrupole moment. In gravitational lens theory, 
the deflection angle \cite{Paez1994,Asada2003} can be used to obtain the lens 
equation, thereby the lensing properties for this new metric with intrinsic 
gravitational quadrupole may be studied. Another application of this metric is 
to calculate the planetary perihelion shift \cite{Yamada2012}. 
These applications will be the aim of forthcoming works.

\appendix
\section{Appendix}

\noindent 
The Ricci tensor components are 

\begin{eqnarray*}
R_{0 0} & = & \frac{1}{4 \rho^2 Y^2 Z^2} \left\{
V Y Z^2 \frac{\partial X}{\partial \theta} \frac{\partial V}{\partial \theta} 
+ V Y^2 Z \frac{\partial X}{\partial r} \frac{\partial V}{\partial r} 
+ 2 \rho^2 Y Z^2 \frac{\partial^2 V}{\partial \theta^2} \right. \\
& - & \left. X Y Z^2 \left[\frac{\partial V}{\partial \theta} \right]^2 
- \rho^2 Z^2 \frac{\partial V}{\partial \theta} 
\frac{\partial Y}{\partial \theta} 
+ \rho^2 Y Z \frac{\partial V}{\partial \theta} 
\frac{\partial Z}{\partial \theta} 
+ 2 \rho^2 Y^2 Z \frac{\partial^2 V}{\partial r^2} \right. \\
& - & \left. X Y^2 Z \left[\frac{\partial V}{\partial r} \right]^2 
+ \rho^2 Y Z \frac{\partial V}{\partial r} \frac{\partial Y}{\partial r} 
- \rho^2 Y^2 \frac{\partial V}{\partial r} \frac{\partial Z}{\partial r}
\right\}
\end{eqnarray*}

\begin{eqnarray*}
R_{0 1} & = & R_{1 0} = 0 \\
R_{0 2} & = & R_{2 0} = 0 \\
R_{0 3} & = & R_{3 0} = \frac{1}{4 \rho^2 Y^2 Z^2} \left\{ 
- 2 W Y Z^2 \frac{\partial X}{\partial \theta} 
\frac{\partial V}{\partial \theta} 
+ V Y Z^2 \frac{\partial X}{\partial \theta} 
\frac{\partial W}{\partial \theta} \right. \\
& - & \left. 
2 W Y^2 Z \frac{\partial X}{\partial r} \frac{\partial V}{\partial r} 
+ V Y^2 Z \frac{\partial X}{\partial r} \frac{\partial W}{\partial r} 
+ X Y Z^2 \frac{\partial V}{\partial \theta} \frac{\partial W}{\partial \theta} 
\right. \\
& + & \left. 
X Y^2 Z \frac{\partial V}{\partial r} \frac{\partial W}{\partial r} 
- 2 \rho^2 Y Z^2 \frac{\partial^2 W}{\partial \theta^2} 
+ \rho^2 Z^2 \frac{\partial W}{\partial \theta} 
\frac{\partial Y}{\partial \theta} \right. \\
& - & \left. \rho^2 Y Z \frac{\partial W}{\partial \theta} 
\frac{\partial Z}{\partial \theta} 
- 2 \rho^2 Y^2 Z \frac{\partial^2 W}{\partial r^2} 
- \rho^2 Y Z \frac{\partial W}{\partial r} \frac{\partial Y}{\partial r}
\right. \\ 
& + & \left. 
\rho^2 Y^2 \frac{\partial W}{\partial r} \frac{\partial Z}{\partial r} \right\}
\\
R_{1 1} & = & \frac{1}{4 \rho^4 Y^2 Z} \left\{ 
- \rho^2 V Y Z \frac{\partial X}{\partial \theta} 
\frac{\partial Z}{\partial \theta} 
- 2 \rho^2 V Y^2 Z \frac{\partial^2 X}{\partial r^2} 
- V^2 Y^2 Z \left[\frac{\partial X}{\partial r} \right]^2 \right. \\  
& - & \left. 
4 \rho^2 Y^2 Z \frac{\partial X}{\partial r} \frac{\partial V}{\partial r} 
+ \rho^2 V Y^2 \frac{\partial X}{\partial r} \frac{\partial Z}{\partial r} 
- \rho^2 X Y Z \frac{\partial V}{\partial \theta} 
\frac{\partial Z}{\partial \theta} \right. \\
& - & \left. 2 \rho^2 X Y^2 Z \frac{\partial^2 V}{\partial r^2} 
- X^2 Y^2 Z \left[\frac{\partial V}{\partial r} \right]^2 
+ \rho^2 X Y^2 \frac{\partial V}{\partial r} \frac{\partial Z}{\partial r} 
+ \rho^4 Z \frac{\partial Y}{\partial \theta} 
\frac{\partial Z}{\partial \theta} \right. \\ 
& - & \left. 2 \rho^4 Y Z \frac{\partial^2 Y}{\partial r^2} 
+ \rho^4 Z \left[\frac{\partial Y}{\partial r} \right]^2 
+ \rho^4 Y \frac{\partial Y}{\partial r} \frac{\partial Z}{\partial r} 
- 2 \rho^4 Y Z \frac{\partial^2 Z}{\partial \theta^2} \right. \\  
& + & \left. 
\rho^4 Y \left[\frac{\partial Z}{\partial \theta} \right]^2 \right\} \\
R_{1 2} & = & R_{2 1} = \frac{1}{4 \rho^4 Y Z} \left\{ 
- 2 \rho^2 V Y Z \frac{\partial^2 X}{\partial \theta \partial r}
- V^2 Y Z \frac{\partial X}{\partial \theta} \frac{\partial X}{\partial r} 
- 2 \rho^2 Y Z \frac{\partial X}{\partial \theta} \frac{\partial V}{\partial r} 
\right. \\
& + & \left. 
\rho^2 V Z \frac{\partial X}{\partial \theta} \frac{\partial Y}{\partial r} 
- 2 \rho^2 Y Z \frac{\partial X}{\partial r} \frac{\partial V}{\partial \theta} 
+ \rho^2 V Y \frac{\partial X}{\partial r} \frac{\partial Z}{\partial \theta} 
- 2 \rho^2 X Y Z \frac{\partial^2 V}{\partial \theta \partial r} \right. \\
& - & \left. 
X^2 Y Z \frac{\partial V}{\partial \theta} \frac{\partial V}{\partial r} 
+ \rho^2 X Z \frac{\partial V}{\partial \theta} \frac{\partial Y}{\partial r} 
+ \rho^2 X Y \frac{\partial V}{\partial r} \frac{\partial Z}{\partial \theta}  
\right\} \\
R_{1 3} & = & R_{3 1} = 0
\end{eqnarray*}

\begin{eqnarray*}
R_{2 2} & = & \frac{1}{4 \rho^4 Y Z^2} \left\{ 
- 2 \rho^2 V Y Z^2 \frac{\partial^2 X}{\partial \theta^2} 
- V^2 Y Z^2 \frac{\partial X}{\partial \theta}^2 
- 4 \rho^2 Y Z^2 \frac{\partial X}{\partial \theta} 
\frac{\partial V}{\partial \theta} \right. \\
& + & \left. \rho^2 V Z^2 \frac{\partial X}{\partial \theta} 
\frac{\partial Y}{\partial \theta} 
- \rho^2 V Y Z \frac{\partial X}{\partial r} \frac{\partial Y}{\partial r} 
- 2 \rho^2 X Y Z^2 \frac{\partial^2 V}{\partial \theta^2} \right. \\ 
& - &  \left. X^2 Y Z^2 \left[\frac{\partial V}{\partial \theta} \right]^2 
+ \rho^2 X Z^2 \frac{\partial V}{\partial \theta} 
\frac{\partial Y}{\partial \theta} 
- \rho^2 X Y Z \frac{\partial V}{\partial r} \frac{\partial Y}{\partial r}
\right. \\  
& + & \left. \rho^4 Z \frac{\partial Y}{\partial \theta} 
\frac{\partial Z}{\partial \theta} 
- 2 \rho^4 Y Z \frac{\partial^2 Y}{\partial r^2} 
+ \rho^4 Z \left[\frac{\partial Y}{\partial r} \right]^2 
+ \rho^4 Y \frac{\partial Y}{\partial r} \frac{\partial Z}{\partial r} 
\right. \\
& - & \left. 2 \rho^4 Y Z \frac{\partial^2 Z}{\partial \theta^2} 
+ \rho^4 Y \left[\frac{\partial Z}{\partial \theta} \right]^2 \right\} \\
R_{2 3} & = & R_{3 2} = 0 \\
R_{3 3} & = & \frac{1}{4 \rho^2 Y^2 Z^2} \left\{ 
- 2 \rho^2 Y Z^2 \frac{\partial^2 X}{\partial \theta^2} 
+ V Y Z^2 \left[\frac{\partial X}{\partial \theta} \right]^2 
- X Y Z^2 \frac{\partial X}{\partial \theta} 
\frac{\partial V}{\partial \theta} \right. \\
& + & \left. \rho^2 Z^2 \frac{\partial X}{\partial \theta} 
\frac{\partial Y}{\partial \theta} 
- \rho^2 Y Z \frac{\partial X}{\partial \theta} 
\frac{\partial Z}{\partial \theta} 
- 2 \rho^2 Y^2 Z \frac{\partial^2 X}{\partial r^2} 
+ V Y^2 Z \left[\frac{\partial X}{\partial r} \right]^2 \right. \\
& - & \left. 
X Y^2 Z \frac{\partial X}{\partial r} \frac{\partial V}{\partial r} 
- \rho^2 Y Z \frac{\partial X}{\partial r} \frac{\partial Y}{\partial r} 
+ \rho^2 Y^2 \frac{\partial X}{\partial r} 
\frac{\partial Z}{\partial r} \right\}
\end{eqnarray*}


\noindent
The scalar curvature is

\begin{eqnarray*}
R & = & \frac{1}{2 \rho^2 Y^2 Z^2} \left\{
2 V Y Z^2 \frac{\partial^2 X}{\partial \theta^2} 
+ 3 Y Z^2 \frac{\partial X}{\partial \theta} \frac{\partial V}{\partial \theta} 
- V Z^2 \frac{\partial X}{\partial \theta} \frac{\partial Y}{\partial \theta} 
\right. \\
& + & \left. 
V Y Z \frac{\partial X}{\partial \theta} \frac{\partial Z}{\partial \theta} 
+ 2 V Y^2 Z \frac{\partial^2 X}{\partial r^2} 
+ 3 Y^2 Z \frac{\partial X}{\partial r} \frac{\partial V}{\partial r} 
+ V Y Z \frac{\partial X}{\partial r} \frac{\partial Y}{\partial r} \right. \\ 
& - & \left. V Y^2 \frac{\partial X}{\partial r} \frac{\partial Z}{\partial r} 
+ 2 X Y Z^2 \frac{\partial^2 V}{\partial \theta^2} 
- X Z^2 \frac{\partial V}{\partial \theta} \frac{\partial Y}{\partial \theta} 
+ X Y Z \frac{\partial V}{\partial \theta} \frac{\partial Z}{\partial \theta}
\right. \\ 
& + & \left. 2 X Y^2 Z \frac{\partial^2 V}{\partial r^2} 
+ X Y Z \frac{\partial V}{\partial r} \frac{\partial Y}{\partial r} 
- X Y^2 \frac{\partial V}{\partial r} \frac{\partial Z}{\partial r} 
- \rho^2 Z \frac{\partial Y}{\partial \theta} 
\frac{\partial Z}{\partial \theta} \right. \\
& + & \left. 2 \rho^2 Y Z \frac{\partial^2 Y}{\partial r^2} 
- \rho^2 Z \left[\frac{\partial Y}{\partial r} \right]^2 
- \rho^2 Y \frac{\partial Y}{\partial r} \frac{\partial Z}{\partial r} 
+ 2 \rho^2 Y Z \frac{\partial^2 Z}{\partial \theta^2} \right. \\
& - & \left. 
\rho^2 Y \left[\frac{\partial Z}{\partial \theta} \right]^2 \right\}
\end{eqnarray*}


\begin{thebibliography}{99}

\bibitem{Abramowicz}
Abramowicz, M.~A., Almergren, G.~J.~E., Klu\'zniak, W. \& 
Thampan, A.~V. 2003. 
\newblock Circular geodesics in the Hartle-Thorne metric.
\newblock ArXiv (gr-qc/0312070). 

\bibitem{Asada2003}
Asada, H., Kasai, M. \& Yamamoto, T. 2003
\newblock Separability of rotational effects on a gravitational lens.
\newblock {\it Phys. Rev. D}, {\bf 67}, 043006 (5 pages).

\bibitem{Asada2005}
Asada, H. 2005
\newblock Effects of a deformation of a star on the gravitational lensing.
\newblock {\it \mnras}, {\bf 356}, 1249--1255. 

\bibitem{Berti}
Berti, E., White, F., Maniopoulou, A. \& Bruni, M. 2005
\newblock Rotating neutron stars: an invariant comparison of approximate
and numerical spacetime models.
\newblock {\em \mnras}, {\bf 358}, 923--938. 

\bibitem{Boshkayev}
Boshkayev, K., Quevedo, H. \& Ruffini, R. 2012
\newblock Gravitational field of compact objects in general relativity.
\newblock {\it Phys. Rev. D}, {\bf 86}, 064043 (13 pages). 

\bibitem{Carmeli}
Carmeli, M. 2001
\newblock {\it Classical Fields}.
\newblock World Scientific Publishing.

\bibitem{Castejon-Amenedo1990}
Castejon-Amenedo, J. \& Manko, V.~S. 1990
\newblock Superposition of the Kerr metric with the generalized Erez-Rosen 
solution.
\newblock {\it Phys. Rev. D}, {\bf 41}, 2018--2020.

\bibitem{Chandrasekhar}
Chandrasekhar, S. 2000
\newblock {\it The Mathematical Theory of Black Holes}.
\newblock Oxford.

\bibitem{Doroshkevich1966}
Doroshkevich, A.~G. Zel'dovich, Ya.~B. \& Novikov, I.~D. 1966
\newblock Gravitational collapse of nonsymmetric and rotating masses.
\newblock {\it JETP}, {\bf 22}, 122--130. 

\bibitem{Erez-Rosen}
Erez, G. \& Rosen, N. 1959
\newblock The gravitational field of a particle possessing a multipole moment.
\newblock {\it Bull. Res. Council Israel}, {\bf 8F}, 47--50.

\bibitem{Ernst1968}
Ernst, F.~J. 1968
\newblock New formulation of the axially symmetric gravitational field problem.
\newblock {\it Phys. Rev.}, {\bf 167}, 1175--1177.

\bibitem{Hartle1968}
Hartle, J.~B. \& Thorne, K.~S. 1968
\newblock Slowly Rotating Relativistic Stars. II. 
Models for Neutron Stars and Supermassive Stars
\newblock {\it AJ}, {\bf 153}, 807--834.

\bibitem{Hearn1999}
Hearn, A.~C. 1999
\newblock {\it REDUCE} (User's and Contributed Packages Manual). 
\newblock Konrad-Zuse-Zentrum f\"ur Informationstechnik, Berlin.

\bibitem{Hoenselaers1979}
Hoenselaers, C., Kinnersley, W. \& Xanthopoulos, B.~C. 1979
\newblock Symmetries of the stationary Einstein-Maxwell equations. VI. 
Transformations which generate asymptotically flat spacetimes with arbitrary 
multipole moments.
\newblock {\it \jmp}, {\bf 20}(12), 2530--2536.

\bibitem{Kerr63}
Kerr, R.~P. 1963
\newblock Gravitational field of a spinning mass as an example of algebraically 
special metrics.
\newblock {\it Phys. Rev. Lett.}, {\bf 11}, 237--238. 

\bibitem{Lewis}
Lewis, T. 1932
\newblock Some Special Solutions of the Equations of Axially Symmetric 
Gravitational Fields.
\newblock {\it Proc. Roy. Soc. Lond.}, {\bf A}, 176--192.

\bibitem{Paez1994}
P\'aez, J. \& Frutos, F. 1994
\newblock Astrometry in the Kerr field in PPN approximation.
\newblock {\it Astrophysics and Space Science}, {\bf 214}, 71--87.

\bibitem{Quevedo1986}
Quevedo, H. 1986
\newblock Class of stationary axisymmetric solutions of Einstein's equations 
in empty space.
\newblock {\it Phys. Rev. D}, {\bf 33}, 324--327.

\bibitem{Quevedo1989}
Quevedo, H. 1989
\newblock General static axisymmetric solution of Einstein's vacuum field 
equations in prolate spheroidal coordinates.
\newblock {\it Phys. Rev. D}, {\bf 39}, 2904--2911.

\bibitem{Quevedo1991}
Quevedo, H. \& Mashhoon, B. 1991
\newblock Generalization of Kerr spacetime.
\newblock {\it Phys. Rev. D}, {\bf 43}, 3902--3906.

\bibitem{Soffel1989}
Soffel, M.~H. 1989
\newblock {\em Relativity in Astrometry, Celestial Mechanics and Geodesy 
(Astronomy and Astrophysics Library)}.
\newblock Springer-Verlag.

\bibitem{Soffel1985}
Soffel, M.~H., Schastok, J., Ruder, H. \& Schneider, M. 1985
\newblock Relativistic Astrometry.
\newblock {\it \apss}, {110}, 95--101.

\bibitem{Weinberg1972}
Weinberg, S. 1972
\newblock {\it Gravitation and Cosmology: Principles and Applications of 
the General Theory of Relativity}.
\newblock John Wiley \& Sons, Inc.

\bibitem{Winicor1968}
Winicour, J., Janis, A.~I. \& Newman, E.~T. 1968
\newblock Static, axially symmetric point horizons.
\newblock {\it Phys. Rev.}, {\bf 176}, 1507--1513.

\bibitem{Yamada2012}
Yamada, K. \& Asada, H. 2012
\newblock Post-Newtonian effects of planetary gravity on the perihelion shift.
\newblock {\it \mnras}, {423}, 3540--3544.

\bibitem{Young1969}
Young, J.~H. \& Coulter, C.~A. 1969
\newblock Exact metric for a nonrotating mass with a quadrupole moment.
\newblock {\it Phys. Rev.}, {\bf 184}, 1313--1315. 

\bibitem{Zeldovich}
Zel'dovich, Ya.~B. \& Novikov, I.~D. 2011
\newblock {\it Stars and Relativity}.
\newblock Dover Publications. 

\end{thebibliography}
\end{document}